\documentclass[11pt,sort&compress]{elsarticle}

\makeatletter
  \long\def\pprintMaketitle{\clearpage
  \iflongmktitle\if@twocolumn\let\columnwidth=\textwidth\fi\fi
  \resetTitleCounters
  \def\baselinestretch{1}%
  \printFirstPageNotes
  \begin{center}%
 \thispagestyle{pprintTitle}%
   \def\baselinestretch{1}%
    {\large\bf\@title}\par\vskip5pt
    \normalsize\elsauthors\par\vskip5pt
    \footnotesize\itshape\elsaddress\par\vskip10pt
    \end{center}%
  \gdef\thefootnote{\arabic{footnote}}%
  }
\makeatother

\newcommand\blfootnote[1]{%
  \begingroup
  \renewcommand\thefootnote{}\footnote{#1}%
  \addtocounter{footnote}{-1}%
  \endgroup
}

\journal{}
\usepackage{tikz}
\usepackage{amsmath}
\usepackage{amsfonts}
\usepackage{lipsum}
\usepackage{graphicx}
\usepackage{xspace}
\usepackage{epsf}
\usepackage{setspace}
\usepackage{float}
\usepackage{abstract}

\usepackage[]{siunitx}
\usepackage{microtype}
\usepackage[margin=1in]{geometry}

\usepackage{stmaryrd}
\SetSymbolFont{stmry}{bold}{U}{stmry}{m}{n}

\usepackage[]{xcolor}
\usepackage{bm}
\newcommand{\ve}[1]{\bm{#1}}

\newcommand{\bu}{\ve{u}}

\newcommand{\bA}{\ve{A}}

\newcommand{\bU}{\ve{U}}

\newcommand{\bM}{\ve{M}}

\newcommand{\bH}{\ve{H}}

\newcommand{\bL}{\ve{L}}

\newcommand{\cM}{\mathcal{M}}

\newcommand{\btheta}{\vec{\boldsymbol{\theta}}}

\newcommand{\hP}{\widehat{P}}
\newcommand{\hb}{\widehat{b}}

\newcommand\Rey{\mbox{\text{Re}}\xspace}

\newcommand{\cD}{\mathcal{D}}

\newcommand{\cS}{\mathcal{S}}
\newcommand{\cO}{\mathcal{O}}
\newcommand{\cN}{\mathcal{N}}

\definecolor{lightblue}{rgb}{0.63, 0.74, 0.78}
\definecolor{seagreen}{rgb}{0.18, 0.42, 0.41}
\definecolor{orange}{rgb}{0.85, 0.55, 0.13}
\definecolor{silver}{rgb}{0.69, 0.67, 0.66}
\definecolor{rust}{rgb}{0.72, 0.26, 0.06}
\definecolor{purp}{RGB}{68, 14, 156}

\colorlet{lightrust}{rust!50!white}
\colorlet{lightorange}{orange!25!white}
\colorlet{lightlightblue}{lightblue}
\colorlet{lightsilver}{silver!30!white}
\colorlet{darkorange}{orange!75!black}
\colorlet{darksilver}{silver!65!black}
\colorlet{darklightblue}{lightblue!65!black}
\colorlet{darkrust}{rust!85!black}
\colorlet{darkseagreen}{seagreen!85!black}

\definecolor{RYB1}{rgb}{0.72, 0.26, 0.06}%dark rust
\definecolor{RYB2}{rgb}{0.18, 0.42, 0.41}%dark seagreen
\definecolor{RYB3}{rgb}{0.63, 0.74, 0.78}%blueish
\definecolor{RYB4}{RGB}{251,220,127}
\definecolor{RYB5}{rgb}{0.69, 0.67, 0.66}
\definecolor{RYB6}{rgb}{0.85, 0.55, 0.13}
\definecolor{RYB7}{RGB}{128, 177, 211}

\definecolor{custompurple}{rgb}{0.578125, 0.40234375, 0.73828125}
\definecolor{custompurple1}{rgb}{0.7734375, 0.77734375, 0.87890625}
\definecolor{custompurple2}{rgb}{0.2734375, 0.046875, 0.51171875}

\definecolor{customgray}{rgb}{0.69921875, 0.69921875, 0.69921875}
\definecolor{customgray2}{RGB}{217,217,217}

\definecolor{customred}{rgb}{0.80859375, 0.1484375, 0.15234375}
\definecolor{customred1}{RGB}{197, 58, 50}
\definecolor{customred2}{RGB}{229, 115, 115}
\definecolor{customred3}{RGB}{239, 154, 154}
\definecolor{customred4}{RGB}{255, 205, 210}

\definecolor{customblue}{rgb}{0.12109375, 0.46484375, 0.703125}
\definecolor{customblue1}{RGB}{58,94,158}
\definecolor{customblue2}{RGB}{102,144,190}
\definecolor{customblue3}{RGB}{153,186,213}

\definecolor{customorange1}{RGB}{255,207,160}
\definecolor{customorange2}{RGB}{183,66,16}

\usepackage{hyperref}
\hypersetup{
  colorlinks=true,
}

\usepackage[labelfont=bf,font=small]{caption}

\usepackage{parskip}
\usepackage{algorithm}
\usepackage{algpseudocode}
\usepackage[qm]{qcircuit}
\usepackage{tabularx} 
\usepackage{booktabs}
\usepackage{arydshln}
\usepackage{soul}

\makeatletter
\def\adl@drawiv#1#2#3{%
        \hskip.5\tabcolsep
        \xleaders#3{#2.5\@tempdimb #1{1}#2.5\@tempdimb}%
                #2\z@ plus1fil minus1fil\relax
        \hskip.5\tabcolsep}
\newcommand{\cdashlinelr}[1]{%
  \noalign{\vskip\aboverulesep
           \global\let\@dashdrawstore\adl@draw
           \global\let\adl@draw\adl@drawiv}
  \cdashline{#1}
  \noalign{\global\let\adl@draw\@dashdrawstore
           \vskip\belowrulesep}}
\makeatother

\usepackage[nameinlink]{cleveref}
\crefname{equation}{}{}
\crefname{appendix}{}{}

\begin{document}

\hypersetup{
  linkcolor=darkrust,
  citecolor=seagreen,
  urlcolor=darkrust,
  pdfauthor=author,
}

\begin{frontmatter}

\title{{\large\bfseries Incompressible Navier--Stokes solve on \\ noisy quantum hardware via a hybrid quantum--classical scheme}}

\author[1]{Zhixin Song}
\author[1,2]{Robert Deaton}
\author[2]{Bryan Gard}
\author[3,4,5]{Spencer H.\ Bryngelson}
\ead{shb@gatech.edu}

\address[1]{School of Physics, Georgia Institute of Technology, Atlanta, GA 30332, USA\vspace{-3pt}}
\address[2]{CIPHER, Georgia Tech Research Institute, Atlanta, GA 30332, USA\vspace{-3pt}}
\address[3]{School of Computational Science \& Engineering, Georgia Institute of Technology, Atlanta, GA 30332, USA\vspace{-3pt}}
\address[4]{Daniel Guggenheim School of Aerospace Engineering, Georgia Institute of Technology, Atlanta, GA 30332, USA\vspace{-3pt}}
\address[5]{George W.\ Woodruff School of Mechanical Engineering, Georgia Institute of Technology, Atlanta, GA 30332, USA}

\date{}
\end{frontmatter}

\begin{abstract}
Partial differential equation solvers are required to solve the Navier--Stokes equations for fluid flow.
Recently, algorithms have been proposed to simulate fluid dynamics on quantum computers.
Fault-tolerant quantum devices might enable exponential speedups over algorithms on classical computers.
However, current and foreseeable quantum hardware introduce noise into computations, requiring algorithms that make judicious use of quantum resources: shallower circuit depths and fewer qubits.
Under these restrictions, variational algorithms are more appropriate and robust.
This work presents a hybrid quantum--classical algorithm for the incompressible Navier--Stokes equations.
A classical device performs nonlinear computations, and a quantum one uses a variational solver for the pressure Poisson equation.
A lid-driven cavity problem benchmarks the method.
We verify the algorithm via noise-free simulation and test it on noisy IBM superconducting quantum hardware.
Results show that high-fidelity results can be achieved via this approach, even on current quantum devices.
Multigrid preconditioning of the Poisson problem helps avoid local minima and reduces resource requirements for the quantum device.
A quantum state readout technique called HTree is used for the first time on a physical problem.
Htree is appropriate for real-valued problems and achieves linear complexity in the qubit count, making the Navier--Stokes solve further tractable on current quantum devices.
We compare the quantum resources required for near-term and fault-tolerant solvers to determine quantum hardware requirements for fluid simulations with complexity improvements.
\end{abstract}

\blfootnote{Code available at \url{https://github.com/comp-physics/NISQ-Quantum-CFD}}

\section{Introduction}\label{s:intro}

With the rapid development of hardware platforms and algorithms~\citep{kim2023evidence, google2023suppressing, moses2023race, bluvstein2024logical, bharti2022noisy}, quantum computing has gained enormous attention due to its potential to solve many large and complex problems faster than classical methods.
The ability of quantum computers to access an exponentially large Hilbert space and exploit unique quantum properties such as superposition and entanglement are at the heart of many current appealing quantum algorithms.
Quantum computing algorithms already promise theoretical speed-up for integer factoring~\citep{shor1999polynomial}, unstructured database search~\citep{grover1996fast}, and many other problems of practical interest~\citep{dalzell2023quantum}.
Numerous quantum PDE solvers have been proposed based on quantum linear system algorithms.
The Harrow--Hassidim--Lloyd (HHL) algorithm~\citep{harrow2009quantum} claims an exponential speed-up for solving linear systems compared to classical iterative methods such as conjugate gradient~\citep{hestenes1952methods}, regardless of the cost of state preparation and readout the quantum solution encoded in complex amplitudes.
HHL and its later improvements~\citep{childs2017quantum, wossnig2018quantum, subacsi2019quantum, an2022quantum} have been successfully adopted to solve various linear PDEs~\citep{cao2013quantum, montanaro2016quantum, berry2017quantum, wang2020quantum, childs2021high, linden2022quantum} and aid in the solution of nonlinear ones~\citep{liu2021efficient, jin2022quantum, lapworth2022hybrid, li2023potential}.

Quantum algorithms for solving computational fluid dynamics (CFD) problems have also gained much attention due to their nonlinear and non-Hermitian nature~\citep{succi2023quantum}.
Some of these works focus on longer-term lattice-based methods, including the lattice Boltzmann method (LBM) and lattice gases~\citep{ljubomir2022quantum,budinski21,yepez2001type,yepez2002quantum,micci2015measurement,kocherla2023fully,kocherla2024two,todorova2020quantum,itani2024quantum,ljubomir2022quantum}, including linearization techniques for handling advective terms~\citep{itani2022analysis,sanavio2024lattice,sanavio2024three} or otherwise treating nonlinearities~\citep{steijl2020quantum}.
\citet{li2023potential} claims a potential quantum advantage for fluid simulation when applying a PDE solver based on quantum linear system algorithms~\citep{berry2017quantum} to Carleman-linearized LBM.
Although they are likewise long-term algorithms, methods based on oracles have also been proposed~\citep{gaitan2020finding} and later developed into quantum circuits for implementation~\citep{gaitan2024circuit}.
\citet{oz2022solving} adopts the quantum PDE algorithm from~\citep{gaitan2020finding} to solve 1D Burgers’ equation and \citet{basu2024quantum} further extend similar approaches to compute dispersal of submarine volcanic tephra.
Perhaps closest to the work presented here is that of \citet{lapworth2022hybrid}, who proposed a hybrid quantum-classical algorithm for solving incompressible Navier--Stokes equations based on the SIMPLE (Semi-Implicit Method for Pressure Linked Equations) algorithm~\citep{pantankar1972calculation} and an HHL linear solver.

The current noisy-intermediate scale quantum (NISQ) hardware capability~\citep{preskill2018quantum} has qubits with meaningfully high error rates.
NISQ methods for solving fluid dynamics problems on actual hardware are limited, and we focus on them here.
Most PDE solves suitable to the NISQ-era use variational strategies~\citep{bharti2022noisy, cerezo2021variational}, including the use of quantum algorithms for pressure Poisson solves in the spectral~\citep{steijl2018parallel,griffin2019investigations} and physical domain~\citep{bharadwaj2023hybrid,lapworth2022hybrid}.

These variational quantum algorithms use shallow parametrized quantum circuits to prepare a general quantum state, combined with classical optimizers and a problem-tailored cost function to (iteratively) prepare the solution.
Among them, the Variational Quantum Linear Solver (VQLS)~\citep{bravo2023variational} is a near-term solution, in some cases, for solving linear systems.
Combined with linearization techniques, one can adopt VQLS to solve simple fluid dynamics problems~\citep{demirdjian2022variational,ali2023performance, liu2024variational}.
To address general nonlinear problems using variational algorithms, \citet{lubasch2020variational} propose the quantum nonlinear processing unit to evaluate nonlinear cost functions given multiple trial states.
Based on that, \citet{jaksch2023variational} develop the variational quantum computational fluid dynamics algorithm to solve the viscous Burgers' equation.

This work proposes a hybrid quantum--classical computing scheme for solving incompressible Navier–Stokes equations on NISQ hardware.
One can combine classical pressure projection methods and variational quantum algorithms to solve the pressure Poisson equation.
Unlike previous approaches that focus on fault-tolerant quantum computers, we perform the calculation on IBM's superconducting quantum processors to solve a 2D flow problem.
Our results show that preconditioning for the linear system can improve the trainability and convergence speed of variational quantum algorithms.
The hybrid algorithm requires a quantum state readout for each simulation time step.

We, in part, address the well-known readout problem~\cite{aaronson2015read}  via a new and efficient quantum state reconstruction method for the real-valued statevector with \textit{linear} time complexity~\citep{htree}.
Conventional methods such as Quantum State Tomography (QST)~\cite{smolin2012efficient} reconstruct the full quantum state as a complex-valued density matrix at the cost of projecting it into an exponential number of Pauli bases.
This improved method instead leverages the fact that many scientific and engineering problems involve only real-valued states.
Then, one separates the readout procedure into amplitude estimation by measuring in the standard computational basis and determining the relative signs between each amplitude. 
This approximate readout method, discussed further in \cref{ss:qst}, simplifies the measurement and can be applied to a broad range of applications beyond PDE solves.
We introduce quantum decoherence noise into the hybrid algorithm for the time-dependent PDE solve.
This is a critical component for designing NISQ-type quantum CFD algorithms and has seen little attention in the literature~\citep{pool2024nonlinear,bharadwaj2024compact}.

This paper begins with \cref{s:qcfd}, reviewing classical CFD solvers for incompressible Navier--Stokes equations, including Chorin's projection method~\citep{chorin1968numerical} and the SIMPLE algorithm.
We introduce the hybrid quantum--classical scheme for incompressible fluid simulations on NISQ devices in \cref{s:hybrid}.
\Cref{s:results} summarizes the simulation and hardware results for a 2D lid-driven cavity flow.
We discuss quantum decoherence noise modeling and the error mitigation and suppression methods we adopt in our hybrid approach.
We further conduct resource estimation to analyze the hardware requirement for solving the same problem with the HHL algorithm.
\Cref{s:conclusions} discusses the presented method's limitations and possible extensions.

\section{Background}\label{s:qcfd}

\subsection{The incompressible Navier--Stokes equations}\label{ss:nse}

The incompressible Navier--Stokes equations are a set of coupled nonlinear PDEs that describe the conservation of mass and momentum.
In dimensionless form, the mass conservation equation is
\begin{gather}
    \nabla \cdot \bu = 0,
    \label{e:continuity}
\end{gather}
and the momentum conservation equation is
\begin{gather}
    \frac{\partial \bu}{\partial t}+(\bu \cdot \nabla) \bu= -\frac{1}{\rho} \nabla p+ \frac{1}{\Rey} \nabla^2 \bu,
    \label{e:momentum}
\end{gather}
where $\bu=(u,v)$ is the flow velocity vector, $\rho$ is the density, $p$ is the pressure, and $\Rey$ is the Reynolds number (ratio of inertial to viscous effects).

\subsection{Pressure projection methods for solving incompressible flow}\label{ss:projection}

The main difficulty of solving the incompressible Navier--Stokes equations is often the coupling of velocity and pressure under incompressibility constraint \cref{e:continuity}.
To transform the governing equations and avoid the pressure-velocity coupling, derived quantities such as the stream function and vorticity could be used instead of the primitive variables $\{u,v,p\}$.
However, this approach poses difficulties in geometry modeling and setting the boundary conditions for 3D problems.

Two numerical schemes for solving viscous incompressible flow equations are pressure-based projection methods and artificial compressibility~\citep{kwak2005computational}.
The former approach decouples the velocity and pressure calculation using a fractional step or pseudo-time-stepping.
Chorin's projection method~\citep{chorin1968numerical} is a well-known pressure projection method, which follows as 
\begin{gather}
    \frac{\bu^{n+1}-\bu^n}{\Delta t}=-\frac{1}{\rho} \nabla p^{n+1}-\left(\bu^n \cdot \nabla\right)\bu^n +\Rey \nabla^2 \bu^n,
    \label{e:projection-discretization}
\end{gather}
with explicit Euler time discretization and $\bu^n$ the velocity at $n^{\text{th}}$ time step.
The time-step $\Delta t$ is chosen following the CFL criterion: $\Delta t \left(\sum_{k} \|\boldsymbol{u}\|_{\text{max}}/ \Delta x_k\right) \leq 1$, where $\Delta x_k$ is the grid spacing in each spatial dimension $k$.
The first step is solving for an intermediate velocity ${\bu}^{*}$ using the discretized momentum equation without the pressure gradient term ($\nabla p^{n+1} = 0$):
\begin{gather}
    \frac{\bu^*-\bu^n}{\Delta t} = -\left(\bu^n \cdot \nabla\right) \bu^n + \Rey \nabla^2 \bu^n.
    \label{e:projection-momentum}
\end{gather}
This velocity field ${\bu}^{*}$ is usually not divergence-free.

\begin{figure}[t]
    \centering
    \includegraphics[]{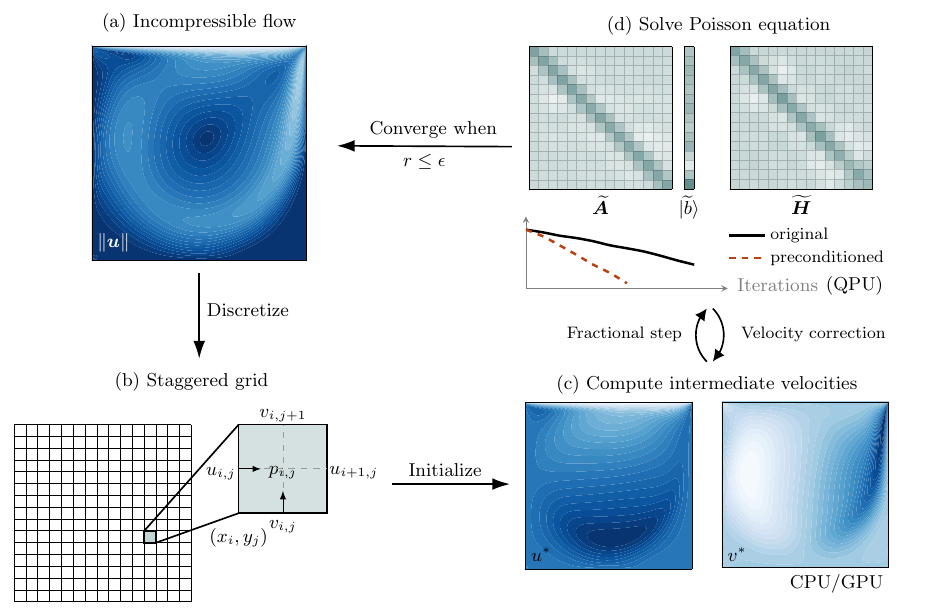}
    \caption{
        We show a schematic of the hybrid quantum--classical scheme for solving incompressible flow problems.
        (a) Velocity contour of the 2D lid-driven cavity flow at $\Rey = 1000$.
        (b) A staggered grid allays the well-known checkerboard pressure problem.
        The pressure is stored at the cell center, and the velocities are stored at the cell interfaces.
        (c) Calculate intermediate velocities and proceed to the classical computer's next fractional time step.
        (d) Solve the Poisson equation with proper preconditioning on a quantum computer and obtain velocity corrections.
        We iterate until the convergence criterion reaches the threshold $\epsilon$.
    }
    \label{f:overview}
\end{figure}
In the second step, the pressure Poisson equation is derived by imposing incompressibility constraint \cref{e:continuity} on the velocity at the next time-step
\begin{gather}
    \nabla^2 p^{n+1}=\frac{\rho}{\Delta t} \nabla \cdot \bu^*.
    \label{e:projection-ppe}
\end{gather}
This projection step is achieved by solving the Poisson equation and is often more computationally expensive than other steps.
Once we solve the Poisson problem, we correct the velocity from $\bu^{*}$ to the next time step as
\begin{gather}
    \bu^{n+1} = \bu^* - \frac{\Delta t}{\rho} \nabla p^{n+1}.
\end{gather}
We iterate the above steps until the residual $r$ converges to a tolerance threshold, here chosen to be $\epsilon= 10^{-5}$, so our results are insensitive to this particular choice.
Residual measures how far the current solution is from satisfying the discrete form of the governing equations and is widely used as a convergence criterion in CFD.
For steady-state problems, one could also monitor the root mean square difference between two consecutive time steps
\begin{gather}
    r_{\text{RMS}} = \frac{\sqrt{\sum_{ij} |\phi_{i,j}^{n+1} - \phi_{i,j}^{n}|^2}}{\sqrt{\sum_{ij} |\phi_{i,j}^{n}|^2}},
    \label{e:rms-residual}
\end{gather}
where the monitored physical quantity $\phi$ are the velocities $(u,v)$ or the pressure $p$.

Different pressure projection schemes exist, such as the SIMPLE (Semi-Implicit Method for Pressure-Linked Equations) algorithm~\citep{pantankar1972calculation}.
Compared to Chorin's projection method, the SIMPLE algorithm uses semi-implicit time-stepping and the velocity correction in \cref{e:projection-momentum} accounts for the pressure solution from the last step.
Both methods can be integrated into our proposed hybrid quantum--classical scheme, see \citet{lapworth2022hybrid} for further discussion.

Another common approach is artificial compressibility or pseudo-compressibility methods~\citep{chorin1997numerical}, where one adds a time-derivative of the pressure term to the continuity equation \cref{e:continuity} as 
\begin{gather}
    \frac{\partial p}{\partial t} + \beta\,\nabla \cdot \bu = 0,
    \label{e:pseudo-compress}
\end{gather}
where $\beta$ is the artificial compressibility.
Such a formulation relaxes the strict requirement to satisfy mass conservation in each step, and the implicit schemes developed for compressible flows can be directly implemented.
However, this method is computationally expensive for transient problems due to dual time-stepping since the pressure field has to go through one complete steady-state iteration cycle in each time step.

\subsection{Quantum solutions for the pressure Poisson equation}\label{ss:q-ppe}

This study focuses on pressure projection schemes.
We transform the pressure Poisson equation into a large-scale system of linear algebraic equations with spatial discretization
\begin{gather}
    \bA |p\rangle = |b\rangle,
    \label{e:ppe-linear}
\end{gather}
where $\bA$ is a sparse Laplacian matrix that only depends on boundary conditions and discrete formats (finite difference or finite volume method), $|p\rangle$ is a vector composed of pressures on all discrete grids, and $|b\rangle$ is a vector containing intermediate velocity divergences and boundary conditions.
We assume the mesh is Cartesian when using the staggered grid in \cref{f:overview}~(b), and one can further extend our method by adopting Rhie--Chow interpolation~\citep{rhie1983numerical}.

Among the family of variational quantum algorithms, the Variational Quantum Eigensolver (VQE)~\citep{tilly2022variational} and Variational Quantum Linear Solver (VQLS)~\citep{bravo2023variational}
are promising candidates for solving linear systems on NISQ devices.
These two approaches both solve the quantum linear system problem $\bA|x\rangle = |b\rangle$ by encoding the solution as the ground-state $|\psi_g\rangle =|x\rangle$ of a problem Hamiltonian $\bH$.
VQE was created to determine the ground-state energy of chemistry molecules based on the Rayleigh--Ritz variational principle~\citep{peruzzo2014variational} and later adopted to solve Poisson equations~\citep{liu2021variational, sato2021variational}.
Given the Hamiltonian mapped into the Pauli basis 
\begin{gather}
    \bH = \sum_{l=1}^{L_h} \alpha_l \bH_l, 
    \label{e:lcu}
\end{gather}
where each Hermitian matrix $\bH_l$ is decomposed in the format of $N$-qubit Pauli strings  $\bH_l\in \left\langle P_1\otimes P_2\cdots\otimes P_N: P_\ell \in\{I, X, Y, Z\}\right \rangle$ and each coefficient $\alpha_l$ is a real number, VQE iteratively prepares the solution with a parameterized quantum circuit (PQC) $U(\btheta)$ by minimizing the expectation value $\sum_l \alpha_l \langle 0|U^\dagger(\btheta) \bH_l U(\btheta)|0\rangle = \sum_l \alpha_l \langle \bH_l \rangle$ as the energy measure.

\begin{figure}
    \centering
    \includegraphics[scale=1]{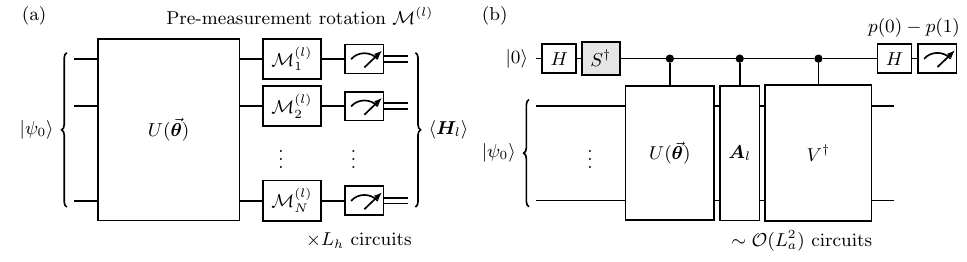}
    \caption{
        Quantum circuit structure for the (a) VQE and (b) VQLS algorithms.
    }
    \label{f:vqe-vqls-circuit}
\end{figure}

\Cref{f:vqe-vqls-circuit}~(a) illustrates the VQE quantum circuit to measure a single term $\langle \bH_l \rangle$.
To compute the total energy inside each iteration, VQE executes $L_h$ circuits.
Thus, the total runtime can be estimated as $t_{\text{VQE}} = N_{\text{itr.}} L_h t_{\text{circ.}}$, where $N_{\text{itr.}}$ is the number of iterations for VQE to reach ground state and $t_{\text{circ.}}$ is the hardware runtime for each circuit.
The pre-measurement rotation gates in Pauli basis $\mathcal{M}^{(l)}\in \{H, H S^\dagger \}$ change the measurement basis for each qubit from computational $Z$-basis to $X$ and $Y$, respectively. 
$H$ (Hadamard gate) and $S^\dagger$ ($-\pi/2$ phase gate) are two basic quantum gate operations.
For example, the pre-measurement rotation gate sequence reads as $\mathcal{M}_1^{(l)}\otimes \mathcal{M}_2^{(l)} \otimes \mathcal{M}_3^{(l)}= I\otimes H \otimes H S^\dagger$ given $\bH_l=Z\otimes X \otimes Y$.
Notice these pre-measurement rotations only consist of single-qubit gates when $\bH_l$ is decomposed in the Pauli basis. 
Decomposing $\bH_l$ into another basis, such as $\{\sigma_+, \sigma_-\}$, where $\sigma_\pm = (X\pm iY)/2$, could lead to a more efficient decomposition such that $L_h = \cO(\mathrm{poly}(N))$ comparing to the worst case scenario for Pauli decomposition of a dense matrix $L_h = \cO(4^N)$. 
Consequently, the pre-measurement rotation circuit is more complex and contains multi-qubit gates~\citep{liu2021variational}. 
Last, one computes the expectation value $\langle \bH_l \rangle$ by measuring in the standard $Z$-basis and post-process the probabilities
\begin{gather}
    \langle \bH_l \rangle = \sum_{z \in\{0,1\}^N}(-1)^{w(z)} p(z),
    \label{e:vqe-expectation}
\end{gather}
where $p(z)$ is the probability of being in the state $|z\rangle = |z_1z_2\cdots z_N\rangle$ with $z_i\in\{0,1\}$ and $w(z)$ is the Hamming weight of the binary string $z$.

VQLS inherits the same variational principle, though includes explicit construction of a $N$-qubit circuit $V$ to load $|b\rangle = V|0\rangle^{\otimes N}$ and additional quantum subroutines such as the Hadamard test to estimate the cost function $C(\btheta) = 1 - |\langle b|\Psi\rangle|^2$, where $|\Psi\rangle = \bA|\psi\rangle/\sqrt{\langle \psi| \bA^{\dagger} \bA| \psi \rangle}$ and $|\psi\rangle = U(\btheta)|0\rangle^{\otimes N} $ is the quantum iterative solution prepared using the same parameterized quantum circuit $U(\btheta)$ as in VQE. 
Here, $|\langle b|\Psi\rangle|^2$ is the cosine similarity between two vectors $|b\rangle$ and $|\Psi\rangle$.
Hence, the cost function has an operational meaning similar to the absolute residual $\||b\rangle-\bA|\psi\rangle\|$.
It is assumed that matrix $\boldsymbol{A}$ is decomposed into a linear combination of unitaries (LCU) $\bA =\sum_{l=1}^{L_a} c_l \bA_l$, where each $\bA_l$ is a unitary matrix and $c_l$ is a complex number.
The VQLS cost function follows as
\begin{gather}
    C(\btheta) = 1 - \frac{\sum_{l, l'} c_l c_{l'}^*\langle 0| U^{\dagger}(\btheta) \bA_{l'}^{\dagger} V|0\rangle\langle 0| V^{\dagger} \bA_l U(\btheta)|0\rangle}{\sum_{l, l'} c_l c_{l'}^* \langle 0| U^{\dagger}(\btheta) \bA_{l^{\prime}}^{\dagger} \bA_l U(\btheta)|0\rangle}.
    \label{e:vqls-cost}
\end{gather}
Last, one repeats the Hadamard test $\cO(L_a^2)$ times to compute the real and imaginary parts of the above terms.

The VQLS circuit for computing a single numerator term $\operatorname{Re}\langle 0| V^{\dagger} \bA_l U(\btheta)|0\rangle$ is illustrated in \cref{f:vqe-vqls-circuit}~(b) given the initial state $|\psi_0\rangle=|0\rangle^{\otimes N}$. 
This circuit represents the Hadamard test using one ancilla qubit initialized in $|0\rangle$ and a controlled unitary sandwiched with two Hadamard gates.
Then, one can calculate $\operatorname{Re}\langle \psi_0| V^{\dagger} \bA_l U(\btheta)|\psi_0\rangle = p(0) - p(1)$, where $p(0)$ and $p(1)$ are the probabilities of measuring $0$ and $1$ in the ancilla qubit.
The shaded extra $S^\dagger$ gate appears only when estimating imaginary parts.
To evaluate denominator terms $\langle 0| U^{\dagger}(\btheta) \bA_{l^{\prime}}^{\dagger} \bA_l U(\btheta)|0\rangle$ in \cref{e:vqls-cost}, one needs to modify the circuit by applying $U(\btheta)$ in the non-controlled way and replacing $V^\dagger$ circuit block with $\bA^\dagger_{l'}$.  
Although VQLS presents as a natural solver for \cref{e:ppe-linear}, it requires meaningfully more quantum resources than VQE on NISQ devices (due to extra controlled gates in Hadamard test) for general engineering problems~\citep{ali2023performance}.
For example, the VQLS circuit runtime for our problem of interest exceeds the coherence time of NISQ hardware.
\Cref{ss:resource} provide a detailed quantum resource estimation comparison between these variation techniques.

\section{Hybrid quantum--classical Scheme}\label{s:hybrid}

\subsection{Overview}\label{ss:overview}

We propose a hybrid quantum--classical scheme for solving the incompressible Navier--Stokes problem.
This scheme combines the projection method with the VQE linear system algorithm.
\Cref{f:overview} illustrates our workflow.
We compute the intermediate velocity field ${\bu}^{*}$ on a classical computer.
We form the discretized Poisson equation and encode the solution as the ground state of an effective Hamiltonian $\bH$ as
\begin{gather}
    \bH = \bA^\dagger(I - |\hb\rangle\langle \hb|)\bA,
    \label{e:ppe-hamiltonian}
\end{gather}
where $\bA^\dagger$ is the Hermitian transpose of the $\bA$ operator and $|\hb\rangle = |b\rangle/\|b\|$.
We use a parameterized quantum circuit $U(\btheta)$ to iteratively prepare the ground state of this Hamiltonian $|\psi(\btheta)\rangle = U(\btheta)|\psi_0\rangle$. 
By optimizing over the parameters $\btheta = (\theta_1, \dots, \theta_m)$ to minimize the cost function
\begin{gather}
    \operatorname*{arg\,min}_{\btheta} \, C(\btheta) = \operatorname*{arg\,min}_{\btheta} \, \langle \psi(\btheta)|\bH|\psi(\btheta)\rangle,
    \label{e:vqe-cost}
\end{gather}
we obtain an approximate solution $|\psi(\btheta_{\text{opt.}})\rangle \approx |\psi_g\rangle$ when $C(\btheta_{\text{opt.}})\leq \gamma$.
$\gamma$ is the termination condition for the optimization loop similar to the residual tolerance in classical iterative linear solvers.
The other common termination condition for VQE is specifying the maximum number of iterations.

To retrieve the pressure solution $|p\rangle$, we require a normalization factor $\cN_p$ to scale the quantum solution such that $\cN_p |\psi(\btheta_{\text{opt.}})\rangle \approx |p\rangle$.
The normalization is computed via the ratio of the largest element of $\bA|\psi(\btheta_{\text{opt.}})\rangle$ and $|\hb\rangle$.
The cost function $C(\btheta)$ has an operational meaning of distance measure between the exact and prepared solutions.
One can show the following lower bound holds in general~\citep{bravo2023variational},
\begin{gather}
    C(\btheta) \geq  \frac{1}{4\kappa^2} \operatorname{Tr}^2 \bigg[ \big|\psi(\btheta_{\text{opt.}})\big\rangle\big\langle \psi(\btheta_{\text{opt.}})\big|- \big|\psi_g \big\rangle\big\langle \psi_g \big| \bigg],
    \label{e:vqe-cost-lowerbound}
\end{gather}
where $\operatorname{Tr}[\,\cdot\,]$ is the trace distance and $\kappa(\bA):=\|\bA\|\left\|\bA^{-1}\right\|$ is the matrix condition number.
The parametrized circuit $U(\btheta)$ is decomposed as a chain of unitary operators
\begin{gather}
    U(\btheta)=\prod_{\ell=1}^{L_{\text{cir.}}} U_\ell(\theta_\ell)W_\ell,
    \label{e:ansatz}
\end{gather}
where $U_\ell(\theta_\ell)= \exp(-i\theta_\ell/2P_\ell)$ with the Pauli operator $P_\ell\in\{X,Y,Z\}$ and $W_\ell$ denotes a fixed (non-parameterized) operator such as two-qubit gates that provide entanglement.

\subsection{Preconditioning the linear system}\label{ss:precondition}

Preconditioning techniques are widely used in iterative sparse linear system solvers.
The common practice is finding a preconditioner $\bM^{-1}$ and applying it to the original linear system as
\begin{gather}
    \underbrace{\vphantom{ \left(\bM^{-1}|b\rangle\right) } \bM^{-1}\bA}_{\widetilde{\bA}} |x\rangle = \underbrace{ \vphantom{ \left(\bM^{-1}|b\rangle\right) }\bM^{-1}|b\rangle}_{|\widetilde{b}\rangle},
    \label{e:precondition}
\end{gather}
such that the matrix condition number of the newly assembled matrix is well behaved $\kappa(\widetilde{\bA})\ll\kappa(\bA)$.
This technique reduces computational complexity since classical sparse linear solvers typically have a $\sqrt{\kappa}$ dependence, compared to the $\kappa^2$ dependence for the HHL algorithm and $\kappa$ dependence for more advanced quantum linear systems algorithms based on discrete adiabatic theorem~\citep{costa2022optimal}.

\citet{clader2013preconditioned} introduced preconditioning to quantum linear system algorithms via a sparse approximate inverse (SPAI) method, where $\bM^{-1}$ is constructed by solving a least-squares procedure for each row of the matrix $\bA$ such that $\|\bM^{-1} \bA-I\|_F^2$ is minimized.
For NISQ linear solvers, \citet{hosaka2023preconditioning} showed that using the incomplete LU factorization as a preconditioning technique can aid VQLS by preparing a higher fidelity solution with fewer iterations in the optimization loop.
The preconditioning involves an LU factorization of $\bA$ (lower and upper triangular): $\bA = \bL \bU$.
One can then drop certain elements in $\bL$ and $\bU$ based on the sparsity pattern of $\bA$ to construct the preconditioner as $\bM = \widetilde{\bL}\widetilde{\bU} \approx \bA$.

\begin{figure}
    \centering
    \includegraphics[]{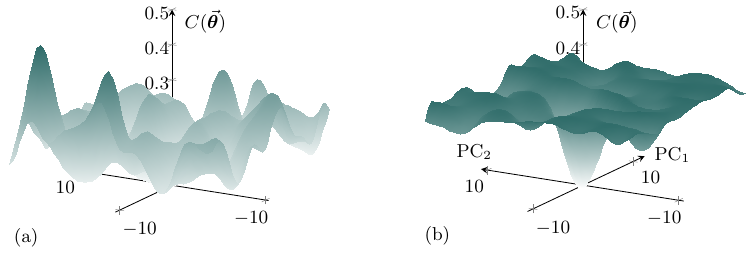}
    \caption{
        Comparison of the (a) original optimization landscape and (b) preconditioned landscape.
        The preconditioned landscape (b) has fewer local minima and thus exhibits better trainability.
        The global minima are located at the center $(0,0)$ in this illustration.
        $\text{PC}_j$ is the $j$-th principal component.
    }
    \label{f:precond-trainability}
\end{figure}

The present hybrid quantum--classical scheme sees an improvement over ILU via a smoothed aggregation algebraic multigrid (AMG) preconditioner.
The AMG used herein constructs a hierarchy $V$-cycle of progressively coarser grids, where smooth error components are associated with low-frequency eigenmodes of matrix $\bA$ that are otherwise challenging to eliminate on the fine grid.
The procedure groups fine-grid points into aggregates to form the coarse grid and then applies a smoother to improve the interpolation between the fine and coarse grids. 
Coarse-level operators are created using Galerkin projection. 
An appropriate multigrid method accelerates the linear solver convergence so long as the operator is elliptic positive definite.
With AMG preconditioning the effective Hamiltonian is
\begin{gather}
    \widetilde{\bH} = \widetilde{\bA}^\dagger(I -|\widetilde{b}\rangle\langle \widetilde{b}|)\widetilde{\bA}.
    \label{e:ppe-hamiltonian-precond}
\end{gather}
Multigrid methods have been recently adopted to variational ansatz design to improve solution quality~\citep{keller2023hierarchical} and enhance trainability~\citep{pool2024nonlinear}.

Here, we use AMG preconditioning to improve the trainability of variational algorithms by reducing the local traps illustrated in \cref{f:precond-trainability}.
In \cref{f:precond-trainability}, the optimization landscape is visualized based on principal component analysis (PCA) of a successful optimization trajectory using the ORQVIZ package~\citep{rudolph2021orqviz}.
The scale of the landscapes is normalized by a factor of $\|\bH\|$ and $\|\widetilde{\bH}\|$ for visualization purposes, but not repeated for the actual simulations.
We only need to marshal the preconditioner once when using Chorin's projection method as $\bA$ is the same for all time steps.
For (semi-)implicit time-stepping tools, such as the SIMPLE algorithm, preconditioning can introduce overhead as $\bM^{-1}$ must be determined separately for each pseudo-time-step.

\subsection{Efficient quantum state readout}\label{ss:qst}

As discussed in \cref{s:intro}, an efficient readout method is key for our (or perhaps any current) hybrid quantum--classical algorithm.
We introduce the Hadamard tree method (HTree) to address this bottleneck~\citep{htree}.
Standard quantum state tomography (QST) reconstructs the density matrix $\rho$ of arbitrary quantum state with complex amplitudes.
So, QST requires collecting $\cO(3^N)$ expectation values by measuring each qubit with the Pauli basis $\{X,Y,Z\}$, where $N$ is the number of qubits involved.
Reconstructing a physical quantum state satisfying $\operatorname{Tr}[\rho]=1$ requires a maximum likelihood post-processing technique.
HTree instead reconstructs real-valued quantum states $|\psi\rangle = \sum_j\psi_j |j\rangle, \psi_j \in \mathbb{R}$ by sampling the magnitude of state amplitude $|\psi_j|$ according to measurement distribution $|\psi_j|^2$.
Hadamard gates on the $(N-k)$-th qubits determine the relative sign between amplitudes $\psi_j$ and $\psi_{j+2^{k}}$.
We compare the runtime- between these two readout methods on real quantum hardware and summarize the results in \cref{t:qst-runtime}.
We use the Qiskit Experiments package to implement QST~\citep{kanazawa2023qiskit}.

\begin{table}[H]
    \centering
    {\small
    \begin{tabular}[t]{l rrrrrr}
        Method & Qubits:\,\, 2 & 3 & 4 & 5 & 6 \\
        \midrule
        QST   & 3 &10  &93 &310 &939 \\
        HTree & 11 &12  &13 &14 &17 \\
        
    \end{tabular}
    }
    \caption{Runtime (seconds) comparison between two readout methods on 27-qubit \texttt{ibmq\_kolkata}.}
    \label{t:qst-runtime}
\end{table}

\begin{figure}
    \centering
    \includegraphics[scale=1]{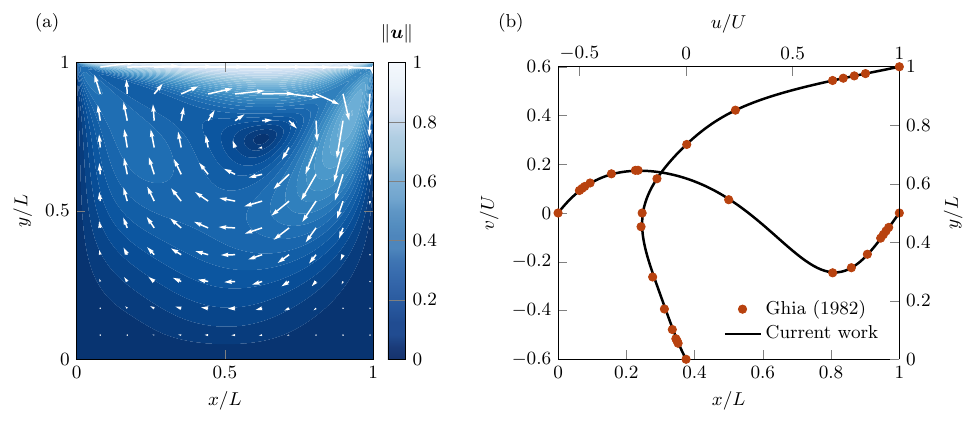}
    \caption{
         Lid-driven cavity flow at $\Rey = 100$ over a $100 \times 100$ grid.
         (a) Velocity magnitude $\|\bu\|= \sqrt{u^2+v^2}$ at $t = 10$.
         (b) Benchmark results and a comparison against a reference solution.
         The horizontal curve is $y$-direction velocity $v$ along the horizontal line through the geometric center of the cavity at $y=0.5$.
         The vertical curve compares the $x$-direction velocity $u$ along the vertical line through the geometric center of the cavity at $x=0.5$ with the reference values of \citet{ghia1982high}.
         The convergence criterion is $r_{\text{RMS}} \leq 10^{-5}$.
        }
    \label{f:cavity-benchmark}
\end{figure}

\section{Results}\label{s:results}

\subsection{CFD benchmark}\label{ss:cavity}

\Cref{f:cavity-benchmark} shows the 2D lid-driven cavity problem.
We consider a square domain with an edge length $L$, and the upper lid moves with velocity $U$ in the horizontal ($x$) direction.
The other three walls are solid and entail a Dirichlet boundary condition on velocity $\bu\big|_{\partial \Omega} = 0$.
The fluid is initially quiescent with $u=v=0$.
We consider a viscous fluid with kinematic viscosity $\nu$ and seek solutions to the incompressible Navier--Stokes equations \cref{e:momentum,e:continuity}.
Solutions are characterized by the Reynolds number $\Rey = UL/\nu$.

We discretize the domain via a uniform staggered grid with $n_k$ grid points in each spatial dimension $k$.
The computational domain thus comprises $n_k^2$ equally spaced points with grid spacing $\Delta x_k = L/(n_k+1)$.
For a scale resolved simulation, the number of grid points $n_k$ can be determined by resolving the Kolmogorov length scale $\eta$ as $L/\eta \sim \cO(\Rey^{3/4})$.
We conduct simulations at $\Rey=100$ and verify the solution against the established benchmark of \citet{ghia1982high}.

\subsection{Simulation results}\label{ss:simulation}

In this section, we conduct numerical simulations that apply the hybrid quantum--classical CFD solver to simulate lid-driven cavity flow.
The parameterized quantum circuits (PQC) $U(\btheta)$, also called the ansatz, used in this study is a specific type of hardware-efficient ansatz (HEA)~\citep{kandala2017hardware}, which consists of the gate set $\cS_{\text{gate}} \in \{R_y(\theta_\ell), \text{CNOT}\}$.
We provide a more detailed analysis to design PQC given hardware resource constraints in \cref{ss:ansatz}.
All simulations, including optimization loops, are conducted using Qiskit~\citep{Qiskit} as an open-source toolkit for quantum computing.

\begin{figure}[t]
    \centering
    \includegraphics[]{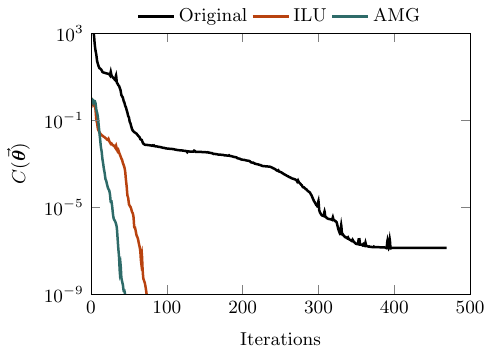}
    \caption{
        Preconditioning helps VQE converge faster to solve the linear system on a $5 \times 5$ mesh.
        The data are averaged from five VQE runs with random initialization.
        The PQC is chosen to be the real-amplitude ansatz with five repetitions.
        The black line shows results without preconditioning.
        ILU is an incomplete LU factorization and AMG is algebraic multigrid.
        The number of iterations reported here could differ from other literature recordings of the number of function evaluations.
        For the L-BFGS-B optimizer, each iteration requires about $30$ evaluations to approximate gradient information.
    }
    \label{f:vqe-convergence-precond}
\end{figure}

We first solve the pressure Poisson problem for $\Rey=100$ at $t=6$ using the Qiskit statevector simulator.
The preconditioning effect is analyzed by repeating the same experiment multiple times with random initial parameters $\theta_\ell \in [-\pi, \pi]$.
We report the convergence history of VQE to prepare the ground state solution with fidelity $F \geq 99\%$ in \cref{f:vqe-convergence-precond}.
The state fidelity $F$ is a cosine distance measure between the true ground-state $|\psi_{g}\rangle$ and prepared state $|\psi(\btheta_{\text{opt.}})\rangle$
\begin{gather}
    F\left(|\psi(\btheta_{\text{opt.}})\rangle, |\psi_{g}\rangle\right) = 
        \left|\langle\psi(\btheta_{\text{opt.}})|\psi_{g}\rangle \right|^2.
    \label{e:state-fidelity}
\end{gather}
Among the different choices of optimizers, we use the L-BFGS-B optimizer~\citep{byrd1995limited}, which we find to consistently outperform other optimizers, like COBYLA and Adam, for statevector simulation.

\Cref{f:vqe-convergence-precond} show that a proper preconditioner can reduce VQE iterations.
We observe the correlation between cost value and other standard distance measure metrics, which include the state fidelity $F$ as $C(\btheta)\approx 1-F$ and the $\ell_2$ norm as $C(\btheta)\approx \||\psi(\btheta_{\text{opt.}})\rangle - |\psi_g\rangle \|_2^2$ when the AMG preconditioner is applied.
The preconditioner aids in the preparation of a higher-fidelity solution by enlarging the spectral gap
\begin{gather}
    \Delta(\bH) = \log_{10} \frac{\big|\lambda_1(\bH)\big|}{\big|\lambda_0(\bH)\big|},
    \label{e:spectral-gap}
\end{gather}
where $\lambda_0$ is the smallest eigenvalue of $\bH$ and $\lambda_1$ is the second smallest eigenvalue.
A large spectral gap means more space for the classical optimizer to explore the solution once $C(\btheta)\leq \lambda_1$.
We further analyze the effect of preconditioners at different mesh sizes in \cref{t:precond}.
The AMG preconditioner consistently creates a larger spectral gap and keeps the matrix condition number low.

\begin{table}[H]
    \centering
    {\small
    \begin{tabular}[t]{crlrrrrrr}
        Mesh & $N_{\text{VQE}}$ & $\Delta t$ & $\kappa(\bA)$ & $\kappa(\widetilde{\bA}_{\text{ILU}})$& $\kappa(\widetilde{\bA}_{\text{AMG}})$ & $\Delta(\bH)$ & $\Delta(\widetilde{\bH}_{\text{ILU}})$ & $\Delta(\widetilde{\bH}_{\text{AMG}})$  \\
        \midrule
        $5\times5$ & 4 & 0.2 & $4.43\times 10^2$ & 8.01 & $\mathbf{1.13}$ & 11.35  & 14.89  & $\mathbf{15.37}$\\
        $9\times9$ & 6 & 0.1 & $3.98\times 10^3$ & 51.51 & $\mathbf{1.33}$ & 10.32  & 13.83  & $\mathbf{15.28}$ \\
        $17\times17$ & 8 & $0.05$ & $3.26\times 10^4$ & $2.94\times 10^2$ & $\mathbf{2.46}$ & 9.97  & 12.17  & $\mathbf{15.26}$ \\
        $33\times33$ & 10 & 0.01 & $2.62\times 10^5$ & $1.53\times 10^3$ & $\mathbf{7.35}$ & 9.12  & 10.75  & $\mathbf{14.89}$ \\
        $65\times65$ & 12 & 0.005 & $2.10\times 10^6$ & $7.56\times 10^3$ & $\mathbf{33.68}$ & 7.70  & 9.89  & $\mathbf{14.21}$ \\
        
    \end{tabular}
    }
    \caption{
        Scaling analysis of two preconditioners with different mesh sizes at $\Rey=100$ and $t=6$, where $N_{\text{VQE}}$ is the number of qubits required for VQE to prepare the solution and $\Delta t$ is the time step size.
    }
    \label{t:precond}
\end{table}

\begin{figure}
    \centering
    \includegraphics[]{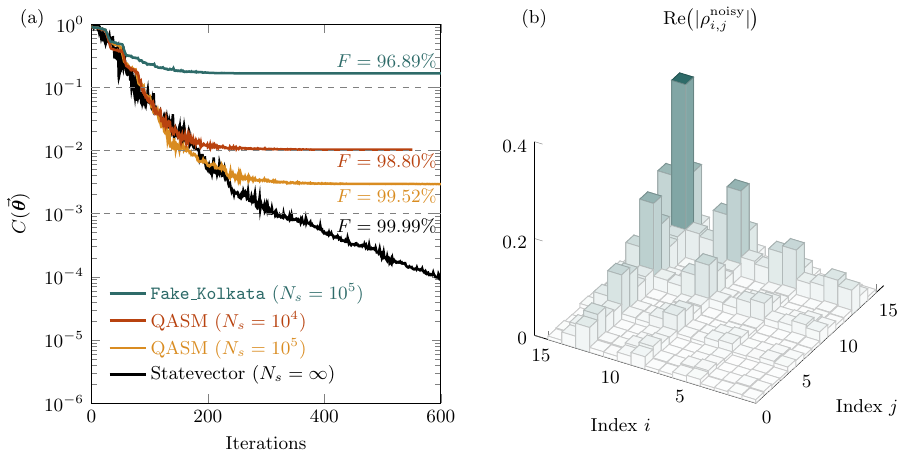}
    \caption{
        (a) Effect of quantum/classical noise on solving the 4-qubit Poisson problem at $t=6$ using VQE.
        (b) Tomography result of the state $\rho_{\text{noisy}} = \cM_{\text{HTree}} \big(|\psi(\btheta_{\text{opt.}})\rangle\big)$ prepared by noisy training on \texttt{Fake\_Kolkata}, where $\cM_{\text{HTree}}(\cdot)$ represent the efficient readout method discussed in \cref{ss:qst}.
    }
    \label{f:sim-comparison}
\end{figure}

Next, we will consider more realistic cases, including finite sampling noise and quantum decoherence noise.
We repeat the same experiment on the Qiskit QASM simulator and switch to the COBYLA optimizer due to its noise resilience performance.
\Cref{f:sim-comparison} shows the convergence history.
In the noise-free simulation, VQE can converge to an optimal cost $C(\btheta_{\text{opt.}})$ of accuracy up to $ 10^{-4}$ and a state fidelity of $99.99\%$ under 700 iterations.
If we let the optimizer keep going until 1600 iterations, it will converge to $C(\btheta_{\text{opt.}}) = 4\times 10^{-8}$ with state infidelity $1-F = 4.14\times 10^{-8}$.
When the state fidelity is higher than $F \geq 99.99\%$, we will report the state infidelity $1-F$ instead.
The sampling-based simulation limits this accuracy to $C(\btheta_{\text{opt.}}) \geq 1/\sqrt{N_s}$, where $N_s$ is the shot number.
Due to the probabilistic nature of quantum computing, one needs to collect many samples to recover the statistical information stored in the quantum state. 
This limitation is common for variational algorithms due to the accuracy required for the prepared ground state, though one can mitigate this via more samples.
We last consider quantum noise from a real quantum device, the 27-qubit \texttt{ibmq\_kolkata} with past calibration data. 
This experiment results in a solution with state fidelity $96.89\%$ compared to the true ground state solution.
We repeat the same noisy VQE experiment multiple times and it typically leads to $3\%\sim5\%$ relative forward errors for solving the linear system at a single time-step.

\subsubsection{Noise modeling on quantum computer}\label{sss:quantum-noise}

Two main categories of quantum noise can be modeled at the circuit level, ignoring all hardware defects: coherent error and decoherence error. 
The former can be caused by miscalibration of the hardware and leads to an undesired over-rotation applied to the circuit $U(\btheta) \rightarrow U(\btheta +\delta\btheta )$, where $\delta\btheta$ sampled from a Gaussian distribution. 
Notably, variational quantum algorithms are insensitive to such coherent errors by design~\cite{cerezo2021variational}. 
Unwanted interactions between the quantum system and the environment cause decoherence errors.
These errors are the most problematic on NISQ hardware and could lead to training issues for VQE. 
Thus, we focus on modeling them in this section.
Decoherence errors can no longer be described as unitary quantum gates and are typically treated as quantum channels in the operator-sum representation~\citep{nielsen2001quantum}
\begin{equation}
    \mathcal{N}(\rho) = \sum_k E_k \rho E^\dagger_k,   
    \label{kraus}
\end{equation}
where $\{E_k\}$ are Kraus operators satisfying the completeness condition $\sum_k E^\dagger_kE_k = I$ and $\rho = |\psi\rangle\langle\psi|$ is the density matrix of a quantum state. 
Typical decoherence errors include the single qubit thermal relaxation error $\mathcal{N}_{\text{TR}}(\cdot)$ describing the effects of energy dissipation ($T_1$ process) and dephasing (associated with $T_2$ process) caused by interaction with the environment at low system temperature 
\begin{gather}
    \mathcal{N}_\text{TR}(\rho) = (1-p_Z-p_{\text{reset}}) \rho + p_Z Z\rho Z^\dagger + p_{\text{reset}} \left( (|0\rangle \langle 0|) \rho (|0\rangle \langle 0|)^\dagger + (|0\rangle \langle 1|) \rho (|0\rangle \langle 1|)^\dagger\right),
    \label{e:noise-thermal-relaxation}
\end{gather}
where $p_Z$ is the probability of dephasing and $p_{\text{reset}}$ stands for the probability of relaxation by projecting to the $|0\rangle$ state.
The other common decoherence error is the $N$-qubit depolarizing channel as a white noise by symmetrically implementing Pauli noises
\begin{equation}
    \mathcal{N}^{(N)}_\text{D}(\rho) = (1-p_\text{D})\rho + \frac{p_\text{D}}{4^N-1} \sum_{i,j,k,\cdots,n=0}^{3} \left(P_i\otimes P_j \otimes \cdots P_n\right)\rho \left(P_i\otimes P_j \otimes \cdots P_n\right)^\dagger,   
    \label{e:noise-depolarizing}
\end{equation}
where $p_\text{D}$ stands for the probability of such an error happening, and each $P_\ell$ is chosen from the Pauli group as $P_0 = I, P_1 = X, P_2 = Y, P_3 = Z$ such that there are $4^N-1$ terms in the summation.

\begin{figure}
    \centering
    \includegraphics[scale=1]{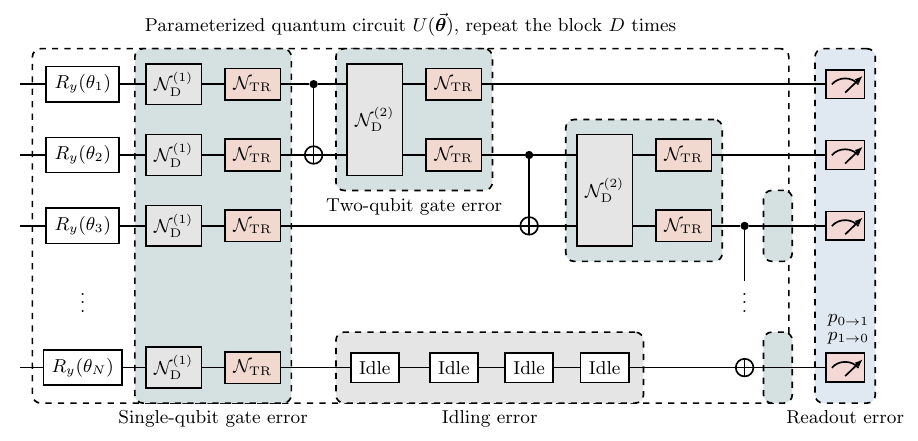}
    \caption{
        Modeling quantum decoherence noise on a generic circuit structure.
        This work does not model the idling error shown in the shaded gray area.
        Executing two-qubit gates adjacent to idling qubits could lead to crosstalk and performance degradation.
        One can adopt dynamical decoupling (DD) as an error suppression technique to tackle idling errors on hardware.
    }
    \label{f:noise-model}
\end{figure}

Given a generic circuit in \cref{f:noise-model}, our noise model consists of single- and two-qubit gate errors and single-qubit readout errors.
The single-qubit gate errors $\mathcal{N}_\text{1q}(\rho) = \mathcal{N}_\text{TR}\circ\mathcal{N}^{(1)}_\text{D}(\rho)$ are composed of a single-qubit depolarizing error channel and a single-qubit thermal relaxation error.
Two-qubit gate errors $\mathcal{N}_\text{2q}(\rho) = (\mathcal{N}_\text{TR}\otimes \mathcal{N}_\text{TR})\circ\mathcal{N}^{(2)}_\text{D}(\rho)$ consist of a two-qubit depolarizing error and a single-qubit thermal relaxation error on both qubits.
Single-qubit readout error $\mathcal{N}_\text{readout}$ flips the classical bit value based on experimentally determined probabilities, Error$_{\text{mea.}} = (p_{0\rightarrow 1} + p_{1\rightarrow 0})/2$.
$p_{0\rightarrow 1}$ stands for the probability of preparing a series of $|0\rangle$ states on hardware but measured as $|1\rangle$.
To determine the probability of errors in $\mathcal{N}_\text{TR}$, we use the gate operation time $t_{\text{g}}$ and hardware $T_1/T_2$ coherence time to estimate $p_{\text{reset}} = 1 - e^{-t_{\text{g}}/T_1}$ and $p_Z = (1-p_{\text{reset}})(1-e^{-t_\text{g}/T_2+t_\text{g}/T_1})/2$.
These data are included in the backend properties through calibrations (see \cref{t:ibm-hardware-metric}) and used to build a fake backend such as \texttt{Fake\_Kolkata} in \cref{f:sim-comparison}.
The probability of the depolarizing error $p_\text{D}$ is set such that the combined average gate infidelity matches the value from backend properties Error$_{\text{1q}}$ (Error$_{\text{2q}}$).

\subsection{Quantum hardware results}\label{ss:hardware}

We study the proposed hybrid quantum--classical scheme on IBM's superconducting quantum hardware.
Considering the noisy nature of quantum hardware, we relax the convergence criterion of the lid-driven cavity flow problem to $r_{\text{max}} \leq 10^{-4}$, where
\begin{gather}
    r_{\text{max}} = \max_{\phi \in \{u,v\}} \left| \phi_{i,j}^{n+1}-\phi_{i,j}^{n} \right|.
    \label{e:max-residual}
\end{gather}
This maximal residual is strictly larger than the root mean square difference of \cref{e:rms-residual}.
We are interested in two questions: (i) What is the minimum tolerable noise to reach the steady-state solution $r_{\text{max}}\leq 10^{-4}$ for the lid-driven cavity benchmark using the proposed hybrid scheme, and (ii) whether current NISQ hardware meets the requirement.
To address (i), we run VQE on a statevector simulator to obtain optimal parameters $\btheta_{\text{opt.}}$ and readout the state $|\psi_{\text{noisy}}\rangle$ from real hardware or noisy simulators using HTree readout method.
One could, in principle, conduct end-to-end hardware experiments by including VQE on-device training.
However, this will introduce extra error sources as the hardware noise behavior drifts with time, and on-device training requires prohibitively long simulation times.

\begin{figure}
    \centering
    \includegraphics[scale=1]{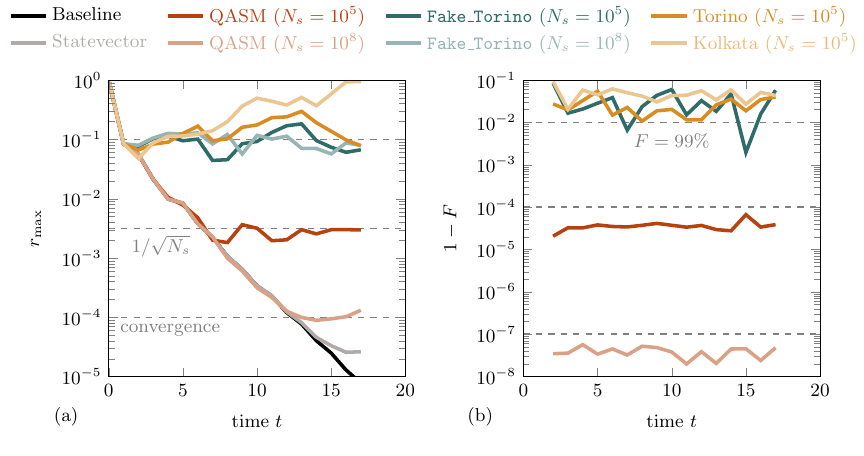}
    \caption{
        (a) Convergence history of the hybrid algorithm on simulators and hardware.
        (b) State infidelity as a quality measure of the hybrid solver in each time step.
    }
    \label{f:ibm-hardware-projection}
\end{figure}

We present the results with a 4-qubit test case in \cref{f:ibm-hardware-projection}.
The calibration data of IBM's quantum hardware is recorded in \cref{t:ibm-hardware-metric}.
The statevector simulation and QASM simulation with a large shot number $N_s=10^8$ meet the loose convergence criterion $r_{\text{max}}\leq 10^{-4}$.
The QASM simulation with $N_s=10^5$ shots shows an accuracy limit near $1/\sqrt{N_s}$.
This conclusion holds for fault-tolerant quantum linear system algorithms, including HHL.
Without quantum noise, the solver still needs many shots to converge.
As a reference value, $10^5$ shots take about 43~seconds on the 133-qubit IBM~Heron processor \texttt{ibm\_torino} and 19~seconds on the 27-qubit Falcon processor \texttt{ibmq\_kolkata}.
The difference is caused by the measurement time $t_{\text{mea.}}$ in two generations of IBM quantum processors as listed in \cref{t:ibm-hardware-metric}.

\begin{table}
    \centering
    \resizebox{\textwidth}{!}{%
    \begin{tabular}[t]{lrrrrrrrr}
    Backend & Qubits & $T_1$ (\unit{\micro\second}) 
    & $T_2$ (\unit{\micro\second}) 
    & Error$_{\text{1q}}$ 
    & Error$_{\text{2q}}$ 
    & Error$_{\text{mea.}}$ 
    & $t_{\text{2q}}$ (\unit{\nano\second}) 
    & $t_{\text{mea.}}$ (\unit{\nano\second}) \\
    \midrule
    \texttt{ibmq\_kolkata} &27 &103 &57 &\num{2.55e-4} & \num{8.93e-3} & \num{1.22e-2} &452  &640\\
    \texttt{ibm\_sherbrooke} &127 &263 &184 &\num{2.17e-4} & \num{7.61e-3} & \num{1.29e-2} &533 &1244 \\
    \texttt{ibm\_torino} &133 &168 &132 &\num{3.08e-4} & \num{3.40e-3} & \num{1.89e-2} &124 &1560
    \end{tabular}
    }
    \caption{
        Calibration data of the IBM quantum computers used in this study.
        The data are averaged among all the qubits.
    }
    \label{t:ibm-hardware-metric}
\end{table}

To address the second question (ii) raised at the beginning of this section, we test the hybrid algorithm on a noisy simulator and hardware with a moderate shot number $N_s=10^5$.
The top two curves in \cref{f:ibm-hardware-projection}~(a) are real hardware results.
\texttt{ibmq\_kolkata} is an older processor with higher two-qubit gate error rates, so we observe an accumulation of error in the maximal residual $r_{\text{max}}$.
While on the latest 133-qubit processor \texttt{ibm\_torino}, we observe a lower error rate but still do not reach convergence.
The noisy simulator results on \texttt{Fake\_Torino} assembles the behavior of real hardware where the detailed noise modeling is described in \cref{sss:quantum-noise}.
The maximal residual consistently stays on the level $r_{\text{max}}\approx 10^{-1}$.
In the presence of quantum noise, increasing the shot number from $N_s=10^5$ to $N_s=10^8$ does not help the hybrid solver converge.
From the perspective of quantum state tomography, the HTree method achieves good fidelity $95\%\sim 99\%$ on real hardware as shown on \cref{f:ibm-hardware-projection}~(b).
We use dynamical decoupling sequences and measurement error mitigation to suppress the effect of quantum noise.
However, the QASM ($N_s=10^8$) simulation result shows that a converging CFD solution requires at least $1-F\leq 10^{-7}$.
The above results indicate that the current NISQ hardware is still too noisy to implement explicit time-stepping numerical schemes for solving time-dependent PDE problems.

\subsection{Resource estimation \& complexity analysis}\label{ss:resource}

This section compares the quantum resources required for three quantum linear systems: VQE, VQLS, and HHL.  
We determine the most suitable one for NISQ hardware based on qubits required, 2-qubit gate counts, circuit depth, and estimated runtime on the 127-qubit IBM Eagle processor \texttt{ibm\_sherbrooke}.
\Cref{t:resource-hhl} summarizes the results.
We further compare the runtime per iteration to classical iterative linear solvers such as conjugate gradient (CG) and GMRES.
This estimation assumes no parallelization for both classical and quantum solvers and does not consider any communication bottlenecks when submitting quantum jobs to IBM's cloud service.
We manually transpile the quantum circuits using the highest optimization level in Qiskit.

As mentioned in \cref{ss:q-ppe}, VQLS would require more quantum resources than VQE on NISQ hardware.
It requires one ancillary qubit to perform the Hadamard test but introduces a deeper circuit mainly due to the controlled application of the parametrized quantum circuit $U(\btheta)$ which composes $54.3\%$ of the estimated runtime on \texttt{ibm\_sherbrooke} and $306$ 2-qubit gates after transpilation.
This leads to each VQLS circuit runtime almost $50\times$ longer than the VQE circuit and beyond the hardware coherence time.
Such an runtime overhead can be mitigated by adopting the Hadamard overlap test as proposed in the original VQLS paper~\citep{bravo2023variational}, where we can significantly reduce circuit runtime at the cost of doubling the qubits used.
The other problematic part is the scaling of number of VQLS circuits $N_{\text{cir.}} = \cO(L_a^2)$ needs to evaluate in each iteration, where $L_a$ is the number of LCU (or Pauli) terms given $\bA =\sum_{l=1}^{L_a} c_l \bA_l$.
Notice the total VQLS runtime for each iteration $N_{\text{cir.}}$ can already be compatible to the total runtime of HHL algorithm.
HHL is known to be a resource-intensive fault-tolerant algorithm which requires much more ancillary qubits, and most of the 2-qubit gate counts come from implementing the $\exp(\bA t)$ time evolution operator.
The results indicate VQLS is only suitable for NISQ hardware when the $m\times m$ matrix $\bA$ has an efficient decomposition such as $L_a = \cO(\text{poly}(\log_2 m))$.
In the original VQLS paper~\citep{bravo2023variational}, the authors consider the Ising-inspired linear system problems satisfying this condition and leading to an empirical scaling of $\cO\left(\left(\log _2(m)\right)^{8.5} \kappa \log (1 / \epsilon)\right)$ through numerical simulations~\citep{patil2022variational}.
Compared to the time complexity of CG for positive definite matrices at each iteration $\cO\left(m s \sqrt{\kappa} \log (1 / \epsilon)\right)$ and GMRES for general matrices $\cO\left(m^2\right)$, VQLS can be advantageous for very large and sparse system $m$.
However, it is unlikely the same scaling holds for general engineering problems through our observation.
From~\cref{t:resource-hhl}, the physical runtime required for VQLS and HHL both exceed the NISQ hardware limitation.
Although obtaining better gate counts using more advanced transpilation methods for fault-tolerant algorithms like HHL is possible, the conclusions here still hold.
We conclude that VQE is the most suitable one for NISQ hardware judging from estimated runtime.

Similar to VQLS, it is difficult to directly give the time complexity of VQE due to its heuristic nature.
As discussed in~\cref{ss:q-ppe}, one can estimate the total runtime of VQE as $t_{\text{VQE}} = N_{\text{itr.}} L_h t_{\text{circ.}}$, given the Hamiltonian in the form of Pauli decomposition $\bH = \sum_{l=1}^{L_h} \alpha_l \bH_l$. 
Since the construction of effective Hamiltonian involves non-trivial $|b\rangle$, $\bH$ is usually a dense Hermitian matrix and leads to approximately $L_h = \cO(m^2)$.
Thus, VQE exhibits no hope to achieve any quantum advantage on solving linear systems over classical iterative solvers unless some novel preconditioner could effectively improve the sparsity of $\bH$ such that $L_h = \cO(\text{poly}(\log_2 m))$ while maintaining a low condition number $\kappa$.
Eventually, we would expect running fault-tolerant quantum linear system algorithms with optimal time complexity $\cO\left(\kappa\log(m/\epsilon)\right)$ on matured hardware with quantum error correction.

\begin{table}
    \centering
    {\small
    \begin{tabular}{llrrrrrr}
        Solver & Regime & Qubits  & 2q gates 
        & Circuit depth  
        & $N_{\text{cir.}}$
        & $t_{\text{cir.}}$ (\unit{\micro\second}) 
        & $t_{\text{total}}$ (\unit{\micro\second}) \\
        \midrule
        CG  &Classical & /  & /  & /  & / & / & 22.2\\
        GMRES &Classical & /  & /  & /  & / & / & 16.8\\
        \midrule
        VQE  &NISQ &4  & 15  & 59  & 136 & \textbf{6.3}  & \textbf{856.8} \\
        VQLS &NISQ &5  & 504 & 2512 & 484 & 291.2 & \num{1.4e5}\\
        HHL  &FT   &12  &\num{3.7e5} & \num{1.7e6} & 1 & \num{2.4e5}  & \num{2.4e5}\\
    \end{tabular}
    }
    \caption{
        Resource estimation for solving a $16\times 16$ linear system on the 127-qubits IBM~Eagle processor \texttt{ibm\_sherbrooke}.
        $t_{\text{cir.}}$ stands for the estimated quantum circuit runtime on this NISQ hardware.
        The coherence time is $T_2 \approx 184$ \unit{\micro\second}.
        The computation returns meaningful results when $t_{\text{cir.}} \leq T_2$.
        $N_{\text{cir.}}$ stands for the number of circuits to evaluate the cost in each iteration.
        Since HHL is not an iterative solver, $N_{\text{cir.}}=1$.
        $t_{\text{total}}$ represents the total runtime on quantum hardware for each iteration of VQE and VQLS.
        For quantum solvers, $t_{\text{total}} \approx N_{\text{cir.}}t_{\text{cir.}}$.
        For classical iterative solver, $t_{\text{total}}$ is averaged among $100$ iterations.
    }
    \label{t:resource-hhl}
\end{table}

\section{Conclusion}\label{s:conclusions}

We propose and test a hybrid quantum--classical algorithm for solving the incompressible Navier--Stokes equations on NISQ hardware.
The algorithm combines the classical pressure projection method and quantum linear system algorithms.
We use an efficient state reconstruction method to address the well-known quantum readout problem, largely unexplored in previous art.
Preconditioning techniques can improve the trainability of variational quantum linear system algorithms.
Although VQLS is widely considered a promising candidate for solving linear systems on NISQ hardware, our resource estimation indicates it would require non-affordable quantum resources unless the matrix $\bA$ has an efficient Pauli or LCU decomposition.
Our results indicate that current quantum hardware is too noisy to solve time-dependent PDE problems with explicit time-stepping.
Besides quantum linear system algorithms, another promising approach for solving PDEs on a quantum computer is via Hamiltonian simulation~\citep{lloyd1996universal} targeting to run on error-corrected quantum hardware and allowing deep quantum circuits.
Compared to our hybrid NISQ algorithm, one clear advantage of Hamiltonian simulation and other fault-tolerant quantum PDE algorithms~\citep{gaitan2020finding,gaitan2024circuit} is they do not require repeated encoding and readout during the computation and hence further control the accumulation of error caused by noisy measurement.
With recent hardware demonstrations for linear PDEs~\citep{sato2024hamiltonian,wright2024noisy}, Hamiltonian simulation could, eventually, provide a different approach for solving the Navier--Stokes equations.

\section*{Declaration of competing interests}

The authors declare no conflicts of interest.

\section*{Data availability}

All code associated with this work is available under the MIT license at \url{https://github.com/comp-physics/NISQ-Quantum-CFD}.

\section*{Acknowledgements}

SHB and BG gratefully acknowledge the support of DARPA under grant no.\ HR0011-23-3-0006. 
This research was developed with funding from the Defense Advanced Research Projects Agency (DARPA).
The views, opinions, and/or findings expressed are those of the author(s) and should not be interpreted as representing the official views or policies of the Department of Defense or the U.S. Government.
The authors thank Nicolas~Renaud for introducing the HTree method, Hsin-Yuan~Huang for quantum tomography methods, Siyuan~Niu for fruitful discussions on dynamical decoupling sequence designs, and Tianyi~Hao for optimization landscape visualization.
SHB acknowledges the resources of the Oak Ridge Leadership Computing Facility, which is a DOE Office of Science User Facility supported under Contract DE-AC05-00OR22725.
IBM Quantum services were used for this work.
The views expressed are those of the authors and do not reflect the official policy or position of IBM or the IBM Quantum team.

\bibliographystyle{bibsty}
\bibliography{ref.bib}

\pagebreak 

\appendix

\section{Quantum circuit and hardware co-design}\label{ss:ansatz}

\begin{figure}[H]
    \text{(a)} \Qcircuit @C=0.7em @R=0.4em{
    &\gate{U}&\ctrl{1}&\qw&\qw&\qw\\
    &\gate{U}&\targ&\ctrl{1}&\qw&\qw\\
    &\gate{U}&\qw&\targ&\ctrl{1}&\qw\\
    &\gate{U}&\qw&\qw&\targ&\qw
    \gategroup{1}{2}{4}{5}{2em}{--}^{\quad\quad ~\times D_1}
    }~
    \text{(b)} \Qcircuit @C=0.7em @R=0.4em{
    &\gate{R_y}&\ctrl{1}&\qw&\qw&\qw\\
    &\gate{R_y}&\targ&\ctrl{1}&\qw&\qw\\
    &\gate{R_y}&\ctrl{1}&\targ&\qw&\qw\\
    &\gate{R_y}&\targ&\qw&\qw&\qw
    \gategroup{1}{2}{4}{5}{2em}{--}^{\quad\quad ~\times D_2}
    }~
    \text{(c)} \Qcircuit @C=0.7em @R=0.4em{
    &\gate{R_y}&\ctrl{1}&\qw&\gate{R_y}&\qw&\qw\\
    &\gate{R_y}&\targ&\qw&\gate{R_y}&\ctrl{1}&\qw\\
    &\gate{R_y}&\ctrl{1}&\qw&\gate{R_y}&\targ&\qw\\
    &\gate{R_y}&\targ&\qw&\gate{R_y}&\qw&\qw
    \gategroup{1}{2}{4}{6}{2em}{--}^{\quad\quad \times D_3}
    }~
    \text{(d)} \Qcircuit @C=0.7em @R=0.4em{
    &\gate{R_y}&\ctrl{1}&\qw&\qw\\
    &\gate{R_y}&\targ&\qw&\qw\\
    &\gate{R_y}&\ctrl{1}&\qw&\qw\\
    &\gate{R_y}&\targ&\qw&\qw
    \gategroup{1}{2}{4}{4}{2em}{--}^{\quad\quad ~\times D_4}
}
    \caption{
        The ansatze considered in this study.
        After repeating $D_j$ times of each block, a final rotation layer enhances expressibility, though it is not illustrated further here.
        (a) HEA uses $U = R_z(\theta_j)R_y(\phi_j)R_z(\varphi_j)$ as a general rotation on the Bloch sphere along with non-parameterized adjacent CNOT gates providing entanglement to prepare the solution state; RAA uses $U=R_y(\theta_j)$ in the rotation layer to reduce the trainable parameters.
        (b) RAA with a pairwise entanglement layer to reduce the circuit duration of one CNOT gate per block.
        (c) Alternating Layered Ansatz (ALT).
        (d) Tensor Network Ansatz (TEN) prepares tensor product states and is expected to have a weaker entangling capability.
    }
    \label{f:circuit-design}
\end{figure}
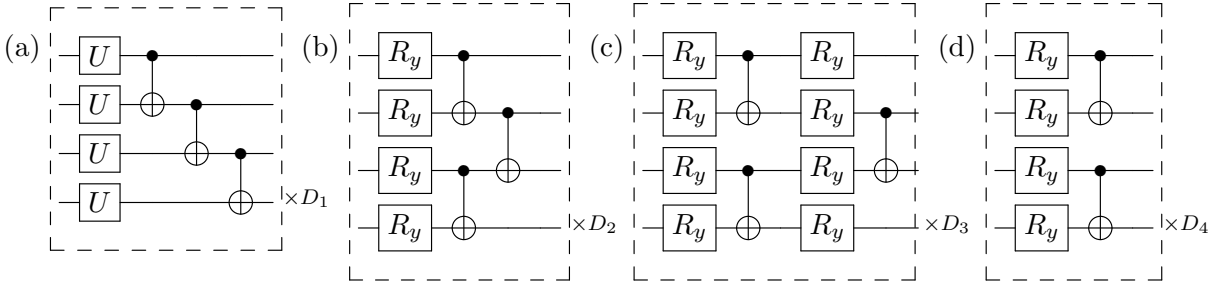

We use different parameterized quantum circuit (PQC) designs for $U(\btheta)$ to prepare the ground-state solution of the pressure Poisson equation on quantum hardware.
We consider four design candidates (illustrated in \cref{f:circuit-design}) widely used in the literature of variational quantum algorithms~\citep{kandala2017hardware, sim2019expressibility, nakaji2021expressibility, cerezo2021cost, araz2022classical} and evaluate them in terms of simulation performance and hardware resource cost.

The Hardware-Efficient Ansatz (HEA)~\citep{kandala2017hardware} was originally proposed for quantum hardware with nearest-neighbor connectivity (such as that of IBM's and Google's superconducting quantum processors) where 2-qubit gates are only applied on adjacent qubits to avoid additional SWAP gates introduced in transpilation~\citep{li2019tackling}.
The Real-Amplitude Ansatz (RAA) is a variant of HEA where the general single qubit rotation layer is limited to $R_y(\theta)$ gate and only generates real-valued states.
Alternating Layered Ansatz (ALT) is composed of blocks forming local 2-designs and proven to be trainable for circuit depth $D\in \cO(\log(N))$ when combined with a local cost function $C_l(\btheta)$~\citep{cerezo2021cost}.
Finally, the Tensor Network Ansatz (TEN) is a family of circuits inspired by classical Tensor Network architectures such as Matrix
Product States (MPS), Tree Tensor Networks (TTN)
and Multi-scale Entanglement Renormalisation Ansatz
(MERA)\citep{araz2022classical}.
For simplicity, we consider a tensor product ansatz here to probe if entanglement is necessary to find the ground state.

\begin{table}[H]
    \centering
    {\small
    \begin{tabular}{lcccrrrr}
        Ansatze & Avg.\ $\bar{F}$ & Expr.\ $\downarrow$ & Ent.\ $\uparrow$  
                & $N_{\text{2q}}$ & $D_{\text{trans.}}$ (DD) 
                &  $t_{\text{circ.}}$ (\unit{\micro\second}) & Coherence \\
        \midrule
        HEA & $99.99\%$  & $\mathbf{0.0003}$ & $\mathbf{0.8149}$ &15 &47 (68) &5.76 & $10.08\%$ \\
        RAA & $99.99\%$  & 0.1899 & 0.7428  &15 &35 (56) &5.76 &$10.08\%$ \\
        RAA-p & $99.27\%$  & 0.1895 & 0.7469  &15 &34 (54) &5.40 & $\mathbf{9.45\%}$ \\
        \cdashlinelr{1-8}
        ALT & $83.82\%$  & 0.2349 & 0.5843 &8 & 29 (34) &2.88 &$5.05\%$ \\
        TEN & $53.04\%$  & 0.6350 & 0.3314 &10 & 29 (30) &2.72 &$4.85\%$ \\
    \end{tabular}
    }
    \caption{
        Summary of a 4-qubit circuit design case study where each PQC candidate contains $24$ trainable parameters except for HEA $N_{\text{param}}=72$.
        The expressibility and entanglement capability are sampled from $10^3$ random samples.
        The average entanglement for the true ground state $|\psi_{g}\rangle$ across different simulation times is $0.46$.
        The transpiled circuit depth with dynamical decoupling (DD) sequence is included.
        Digital DD sequence generally increases transpiled circuit depth but not actual hardware runtime $t_{\text{circ.}}$.
        Coherence is calculated as $t_{\text{circ.}}/T_2 $, where the average qubit lift-time $T_2 \approx 57.15$ \unit{\micro\second} for \texttt{ibmq\_kolkata}.
    }
    \label{t:circuit-performance}
\end{table}

To evaluate each circuit candidate performance, we randomly sample $K=10$ pre-conditioned Hamiltonian $\{ \widetilde{\bH}(t)\}_{i=1}^{K}$ at different times of the simulation with $\Delta t =1$.
We then randomly initiate the parameters of VQE and solve for ground-state preparation with the L-BFGS-B optimizer~\citep{byrd1995limited} on the Qiskit statevector simulator.

\begin{figure}[t]
    \centering
    \includegraphics[scale=1.0]{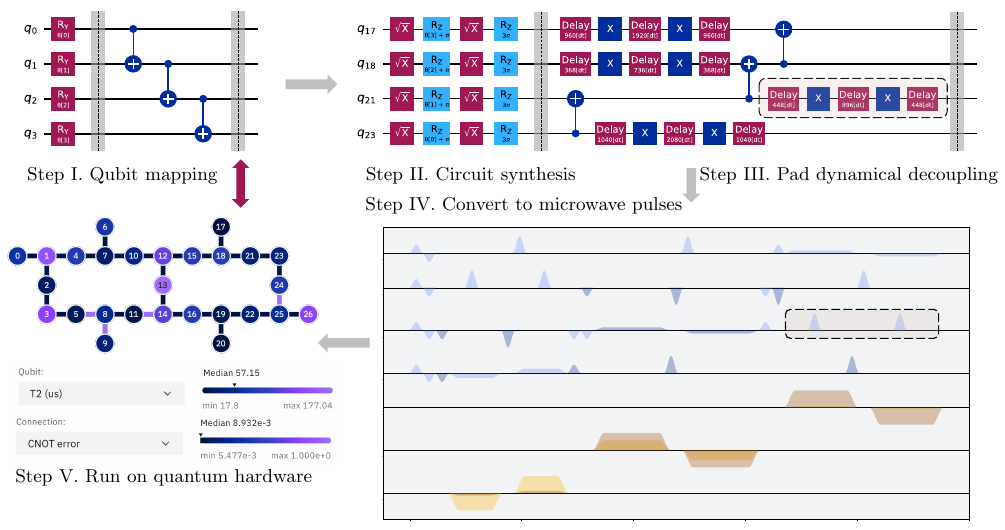}
    \caption{
        Visualization of the transpilation process of 4-qubit RAA ($D_1=1$) on 27-qubit \texttt{ibmq\_kolkata}.
    }
    \label{f:circuit-transpile}
\end{figure}

We calculate the expressibility and the entanglement capability \citep{sim2019expressibility} as auxiliary measures for each ansatz candidate.
The ansatz expressibility describes its capability to sample uniformly from the Hilbert space and approximate arbitrary states.
It can be defined as
\begin{gather}
    \operatorname{Expr} = \cD_{\mathrm{KL}}\left(\hP_{\mathrm{PQC}}(F_{\btheta}) \| P_{\text {Haar }}(F)\right),
    \label{e:expressibility}
\end{gather}
where the Kullback--Leibler (KL) divergence~\cite{kullback1951information} measures the distance between $\hP_{\mathrm{PQC}}(F_{\btheta})$ distribution of state fidelities between two randomly sampled parametrized states, and $P_{\text{Haar}}(F)$ is the state fidelity distribution for the ensemble of Haar random states.
$\operatorname{Expr}$ closer to zero indicates more expressivity for an ansatz.
Following the original definition~\citep{sim2019expressibility}, we calculate the entanglement capability as the average Meyer--Wallach $Q$-measure~\citep{meyer2002global} from an ensemble of randomly sampled states
\begin{equation}
    \mathrm{Ent}
    = \frac{1}{|S|} \sum_{\theta_j \in S} Q\big(\big|\psi(\theta_j)\big\rangle\big)
    = \frac{2}{|S|} \sum_{\theta_j \in S}\left(1-\frac{1}{n}
        \sum_{k=1}^n \operatorname{Tr}\left[\rho_k^2(\theta_j)\right]\right),
\end{equation}
where $\rho_k$ is the reduced density matrix of the $k$-th qubit and $|S|$ represents the size of samples.
Notice $Q=0$ for any product state and $Q=1$ for the GHZ state $|\psi\rangle_{\text{GHZ}} = (|0\rangle^{\otimes n} + |1\rangle^{\otimes n})/\sqrt{2}$.
We further record the number of 2-qubit gates involved $N_{\text{2q}}$, transpiled circuit depth $D_{\text{trans.}}$ and hardware runtime $t_{\text{circ.}}$ on 27-qubit \texttt{ibmq\_kolkata}.
The circuits are transpiled using the highest optimization level and adopt XX dynamical decoupling sequence~\citep{tripathi2022suppression,das2021adapt,niu2022effects, ezzell2023dynamical} to suppress decoherence and crosstalk during the idling time of the qubits, as illustrated in \cref{f:circuit-transpile}.

We report a $N=4$ case study in \cref{t:circuit-performance}.
The random seeds for parameter generation, simulator, and optimizer are fixed to ensure a fair comparison of averaged state fidelity $\bar{F}$.
One can conclude that the real-amplitude ansatz (RAA) outperforms all the other candidates in terms of fidelity and resource burden.
In this case, switching from the linear entanglement structure via RAA pairwise only reduces $0.6\%$ of the coherence time.
The difference can be as high as $5\%$ ($\approx 2.8$ \unit{\micro\second}) at $N=10$ with $D_2=11$.
HEA has higher expressibility and entanglement capability but achieves the same average fidelity with $3$~times more parameters.
The results from TEN emphasize that quantum entanglement is a necessary resource in general state preparation.
To further enhance performance, one can consider strategies such as ADAPT-VQE~\citep{grimsley2019adaptive} to design problem-tailored ansatze,  noise-adaptive circuit design~\citep{wang2022quantumnas} to cope with hardware noise, and pulse-level ansatze~\citep{liang2024napa} to suppress decoherence due to much shorter duration on hardware.

\end{document}